\documentclass[conference]{IEEEtran}
\IEEEoverridecommandlockouts
\usepackage{cite}
\usepackage{amsmath,amssymb,amsfonts}
\usepackage{algorithmic}
\usepackage{graphicx}
\usepackage{textcomp}
\usepackage{xcolor}
\def\BibTeX{{\rm B\kern-.05em{\sc i\kern-.025em b}\kern-.08em
    T\kern-.1667em\lower.7ex\hbox{E}\kern-.125emX}}
\begin{document}

\title{Adaptive Graph Normalized Sign Algorithm
\thanks{Changran Peng and Yi Yan contributed equally. Corresponding author: Ercan E. Kuruoglu. This work is supported by Shenzhen Science and Technology Innovation Commission under Grant JCYJ20220530143002005, Shenzhen Ubiquitous Data Enabling Key Lab under Grant ZDSYS20220527171406015, and Tsinghua Shenzhen International Graduate School Start-up fund under Grant QD2022024C.}
}

\author{
    \IEEEauthorblockN{
        Changran Peng\IEEEauthorrefmark{1}\IEEEauthorrefmark{2}, Yi Yan\IEEEauthorrefmark{1}\IEEEauthorrefmark{2}, Ercan E. Kuruoglu\IEEEauthorrefmark{1}\IEEEauthorrefmark{2}
    }
    \IEEEauthorblockA{\IEEEauthorrefmark{1} Tsinghua-Berkeley Shenzhen Institute, Tsinghua University}
    \IEEEauthorblockA{\IEEEauthorrefmark{2} Institute of Data and Information, Tsinghua Shenzhen International Graduate School}
}
% \author{\IEEEauthorblockN{Anonymous Authors}}

\maketitle

\begin{abstract}
Efficient and robust prediction of graph signals is challenging when the signals are under impulsive noise and have missing data. 
Exploiting graph signal processing (GSP) and leveraging the simplicity of the classical adaptive sign algorithm, we propose an adaptive algorithm on graphs named the Graph Normalized Sign (GNS). 
GNS approximated a normalization term into the update, therefore achieving faster convergence and lower error compared to previous adaptive GSP algorithms. 
In the task of the online prediction of multivariate temperature data under impulsive noise, GNS outputs fast and robust predictions.
\end{abstract}


\section{Introduction}

Graph signal processing (GSP) transfers traditional signal processing concepts to the graph domain through graph shift, graph convolution, graph, and Fourier transform (GFT) \cite{b1}. 
For graph signals that contain the time dimension in applications like wind speed prediction \cite{b2}, brain connectivity analysis \cite{b3}, traffic forecasting \cite{b4} etc., online prediction is preferred when we require efficient and robust results in real-time. 
The Graph Least Mean Squares algorithm (GLMS) \cite{b5} was proposed as a solution to this problem as a form of graph adaptive filters utilizing GSP techniques. 
The Graph Normalized Least Mean Squares algorithm (GNLMS) \cite{b6} introduces a normalization term into GLMS and achieves faster convergence than GLMS. 
However, LMS has the underlying assumption of Gaussian noise.
When the noise is heavy-tailed and impulsive, LMS methods become unstable. 
Therefore, the Graph Least Mean $p^{th}$ algorithm (GLMP) \cite{b7} and its normalized version the Graph Normalized LMP algorithm (GNLMP) \cite{b8} were introduced under the assumption that the noise is symmetric alpha-stable distribution (S$\alpha$S), derived from minimum dispersion criterion. 
The Graph-Sign algorithm (G-Sign) \cite{b9} was developed as an extension of the traditional adaptive sign algorithm to graphs. 
G-Sign removes the need for prior knowledge of noise distribution and achieves time efficiency by solving an $l_1$-norm problem compared to the previous four algorithms.

GNLMS and GNLMP contain spectral domain normalization aimed at reducing the number of iterations needed to converge as well as minimizing the signal and error caused by noise.
Noting that G-Sign achieves better time efficiency without the utilization of normalization, it is worth investigating the effect of normalization introduced to the already efficient G-Sign.
This paper proposes the adaptive graph normalized sign algorithm (GNS). GNS utilizes spectral domain normalization in the update which significantly increases convergence speed and achieves better prediction performance compared to previous adaptive GSP algorithms. 
Experiments on graph-based temperature data prove the effectiveness of GNS.

\section{Background}
The Laplacian matrix $\mathbf{L}$ represents an unweighted and undirected graph with the $ij^{th}$ term being $-1$ for an edge connecting nodes $i$ and $j$; the $i^{th}$ diagonal term is the number of edges of node $i$.
Graph signal $\boldsymbol{x}$ is a function value defined on the nodes. 
GFT is defined based on the eigenvalue decomposition $\mathbf{L}=\mathbf{U\Lambda U}^\mathit{T}$; the orthonormal eigenvector matrix is $\boldsymbol{U}$ and the diagonal eigenvalue matrix is$\Lambda$.
Eigenvalues are ordered in increasing magnitude as an analogy to frequency. 
GFT transforms $\boldsymbol{x}$ from spatial domain to spectral domain: $\boldsymbol{x}$: $\boldsymbol{s} = \mathbf{U}^\mathit{T}\boldsymbol{x}$ and inverse GFT does the opposite: $\boldsymbol{x} = \mathbf{U}\boldsymbol{s}$. 
Missing data in graph signal is modeled by a masking matrix $\mathbf{D}_s$: the $i^{th}$ diagonal is 0 if missing and 1 if observed.

\section{Adaptive Graph Normalized Sign Algorithm}
In adaptive graph filters, the task is to make online predictions from a noisy and missing observation $\boldsymbol{y}[t]=\mathbf{D}_{\mathcal{S}}(\boldsymbol{x}_g[t]+\mathbf{w}[t])$, where $\boldsymbol{x}_g[t]$ is the ground truth.
The noise $\mathbf{w}[t]$ is modeled by the S$\alpha$S distribution, which is a generalization of the Gaussian distribution that can model impulsiveness.
The variable $\boldsymbol{e}[t]=\mathbf{D}_\mathcal{S}(\boldsymbol{y}[t]-\hat{\boldsymbol{x}}[t])$ is the current step estimation error. 
A special case of the minimum dispersion criterion is used to form the cost function of the G-Sign algorithm by $l_1$-norm optimization to obtain a fast and robust estimation of $\boldsymbol{x}_g[t]$ under impulsive noise: 
\begin{equation}
    J(\hat{\boldsymbol{x}}[t])=\mathbb{E}\left\|\boldsymbol{y}[t]-\mathbf{D}_{\mathcal{S}}\mathbf{B}\hat{\boldsymbol{x}}[t]\right\|_1^1,
    \label{cost}
\end{equation}
in which $\mathbf{B} = \mathbf{U \Sigma}\mathbf{U}^T = \mathbf{U}_\mathcal{F}\mathbf{U}_\mathcal{F}^T$ and $\mathbf{\Sigma}$ is a predefined bandlimited filter. 
Using bandlimitedness $\boldsymbol{x} = \mathbf{B}\boldsymbol{x}$, G-Sign algorithm can be derived by stochastic gradient descent:
\begin{equation}
        \hat{\boldsymbol{x}}\left[t+1\right] =\hat{\boldsymbol{x}}[t]-\mu_{s}\frac{\partial  J\left(\hat{\boldsymbol{x}}[t]\right) }{\partial \hat{\boldsymbol{x}}[t]}
    =\hat{\boldsymbol{x}}[t]+\mu_{s}\mathbf{B}Sign(e[t])
    \label{sign_update_1}
\end{equation}
% The $Sign$ is obtained considering the point of discontinuity of the derivative at 0. 

In GNLMS and GNLMP algorithms, a spectral normalization was derived, aiming to accelerate the convergence and reduce the estimation error of GLMS and GLMP.
Noticing this absence in the G-Sign, we propose the GNS algorithm, which has even faster convergence behavior compared to the previously fast G-Sign algorithm, while maintaining robustness under impulsive noise. 
To obtain the normalization, we transform \eqref{sign_update_1} into spectral domain by GFT:
\begin{align}
    \hat{\boldsymbol{s}}\left[t+1\right] =\hat{\boldsymbol{s}}[t]+\mu_{n}\mathbf{M}[t]Sign(\mathbf{U}^T e[t]),
    \label{sign_update_2}
\end{align}
% Following the derivation in GNLMP\cite{b2}, $M[t]$ is solved by setting the derivative to be 0:
where $\hat{\boldsymbol{s}}\left[t\right] = \mathbf{U}_\mathcal{F}^\mathit{T}\hat{\boldsymbol{x}}\left[t\right]$ is the spectral prediction at $t$.
Now, defining \textit{a posteriori error} $\boldsymbol{\varepsilon}[t]=\mathbf{D}_\mathcal{S}\left(\hat{\boldsymbol{x}}[t]-\mathbf{U}_\mathcal{F}\hat{\boldsymbol{s}}\left[t+1\right]\right)$ to be the prediction error between current step and the next step, we can take the derivative of $\boldsymbol{\varepsilon}[t] - \boldsymbol{e[t]}$ with respect to $\mathbf{M}[t]$.
Exploiting the property $\mathbf{I} = \mathbf{U}_\mathcal{F}^\mathit{T}\mathbf{U}_\mathcal{F}$, we obtain  $\mathbf{M}[t]$ for GNS by setting the previous derivative to be 0:
\begin{equation}
    \mathbf{M}[t]=\left(\mathbf{U}_\mathcal{F}^T\mathbf{D}_\mathcal{S}\text{diag}\left(\lvert\boldsymbol{y}[t]-\hat{\boldsymbol{x}}[t]\rvert^{-1}\right)\mathbf{U}_\mathcal{F}\right)^{-1}.
    \label{M}
\end{equation}
Each calculation $\mathbf{M}[t]$ at $t$ will bring extra computational costs. 
Assuming that the estimation $\hat{\boldsymbol{x}}[t]$ resembles the spectrum of the ground truth $\boldsymbol{x}_g[t]$, we further assume that the main cause of estimation error is noise.
Although the S$\alpha$S noise has no analytical form or defined variance, we can use the fractional lower order moment (FLOM) \cite{b10} to model noise statistics.
% \begin{equation}
%     \begin{split}
%     \text{FLOM}(p,\alpha,\gamma)=\mathbb{E}\mathbf{\lvert{X}\rvert}^p=C\left(p,\alpha\right)\gamma^{p/\alpha},\\
%     \text{with } C\left(p,\alpha\right)=\frac{2^{p+1}\Gamma\left(\frac{p+1}{2}\right)\Gamma\left(-\frac{p}{\alpha}\right)}{\alpha\sqrt{\pi}\Gamma\left(-\frac{p}{2}\right)}.
%     \label{FLOM}
%     \end{split}
% \end{equation}
Using these facts, we can approximate $\mathbf{M}[t]$ by 
\begin{equation}
    \mathbf{M}[t] \approx \mathbf{M} = \left(\mathbf{U_F}^T\mathbf{D}_\mathcal{S}\mathbf{RU_F}\right)^{-1},
    \mathbf{R} = (E\lvert\boldsymbol{w}[t]\rvert)^{-1}\mathbf{I}.
\end{equation}
The update function of the GNS algorithm is formalized as
\begin{equation}
        \hat{\boldsymbol{x}}\left[t+1\right]= \hat{\boldsymbol{x}}[t]+\mu_{n}\mathbf{B_n}Sign(\mathbf{D}_{\mathcal{S}}(\boldsymbol{y}[t]-\hat{\boldsymbol{x}}[t])), 
\end{equation}
where $\mathbf{B_n} = \mathbf{U_F}\left(\mathbf{U_F}^T\mathbf{D}_\mathcal{S}\mathbf{RU_F}\right)^{-1}\mathbf{U_F}^T$.

\section{Experiment Results and Discussion}
Our proposed GNS algorithm is tested in two perspectives: convergence speed and prediction accuracy. 
The dataset is a time-varying graph signal of $95$ hourly temperatures collected from $N=197$ weather stations \cite{b11}. 
The latitude and longitude of the weather stations are used to form an $8$-nearest-neighbor graph. 
We follow settings from \cite{b9} of only allowing the signals on $130$ nodes to be observed while the rest are missing.
The bandlimited filter is set to have $|\mathcal{F}| = 120$ frequencies through the greedy approach that maximizes the spectral content \cite{b6}. 
The noise is S$\alpha$S noise with $\gamma = 0.1$ while $\alpha$ varys from 1.05 to 1.25. We compare our GNS algorithm with G-Sign in \cite{b9} and GLMS in \cite{b5} with 1000 experiment runs in terms of Mean Squared Error(MSE). We further compare the convergence speed of GNS, G-Sign, GLMS, and GLMP on a time-invariant graph signal in terms of the number of iterations that reach a steady state.

\begin{table}[htbp]
    \centering
    \caption{Prediction spatial MSE under different $\alpha$ settings.}
    \begin{tabular}{c c c c c c}
        \hline
        $\alpha$  & 1.05 & 1.1 & 1.15 & 1.2 & 1.25  \\
        \hline
        GLMS & 36.6242 & 10.8817 & 4.6772 & 3.0032 & 2.5043 \\
        \hline
        G-Sign & 2.5916 & 2.5916 & 2.6714 & 2.6680 & 2.6515 \\
        \hline
        GNS(ours) & \bf{2.4208} & \bf{2.4205} & \bf{2.4313} & \bf{2.5512} & \bf{2.4554} \\
        \hline
    \end{tabular}
    \label{table_MSE}
\end{table}

The spatial MSE of all methods under different noise settings is shown in TABLE I. We can find that sign-based algorithms are not affected much by the change of $\alpha$, and GNS has better prediction performance than G-Sign under 
all different $\alpha$ settings from 1.05 to 1.25. The convergence speed of all methods under alpha-stable noise when $\alpha = 1.1$ is shown in Fig.1. GNS takes about only half iterations to reach a steady state compared to G-Sign, due to the normalization term. Fig.2. shows the performance comparison of GNS, G-Sign, and GLMS when $\alpha = 1.1$. We can find that under the setting of $\alpha = 1.1$, the noise has strong impulsiveness so GLMS doesn't perform well. Our proposed method performs better than G-Sign in most time steps, which proves its effectiveness.

% \begin{figure}[t]
%     \centering
%     \includegraphics[width=0.6\linewidth, trim={0 0 0 120pt},clip]{converge.png}
%     \caption{Spectral MAE of GLMS, GLMP, G-Sign and GNS.}    
%     \label{fig_convergence}

% \end{figure}

% \begin{figure}[hbtp]
%     \centering
%     \includegraphics[width=0.6\linewidth, trim={0 0 0 140pt},clip]{mse_1.1.png}
%     \caption{Spatial MSE of GNS, G-Sign and GLMS.}    
%     \label{fig_convergence}
% \end{figure}

\begin{figure}[t]
    \centering
    \vspace{-10 pt}
    \includegraphics[width=0.8\linewidth, trim={0 0 0 20pt},clip]{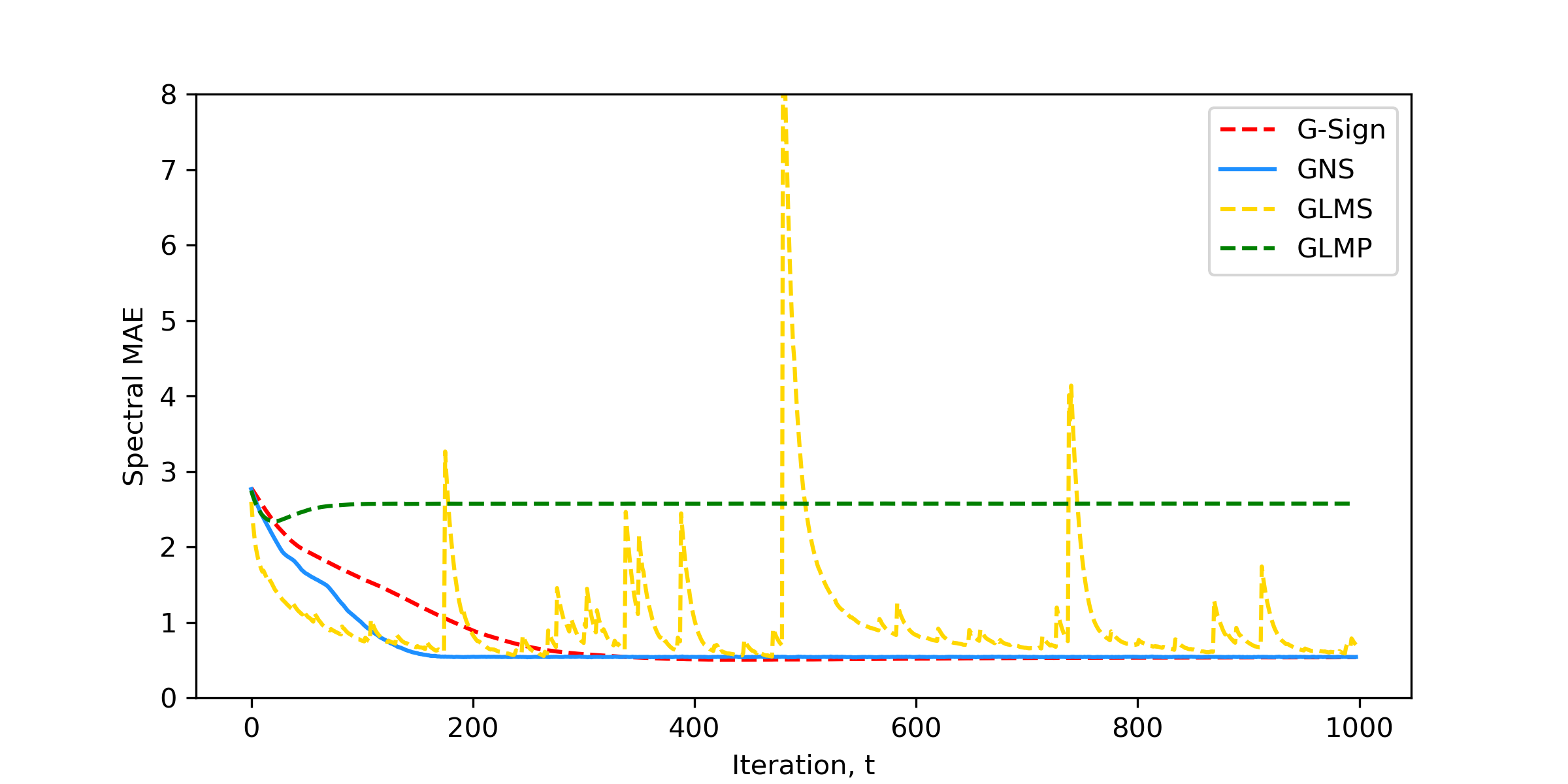}
    \vspace{-10 pt}
    \caption{Spectral MAE of GLMS, GLMP, G-Sign and GNS.}    
    \vspace{-10 pt}
    \label{fig_convergence}
\end{figure}

\begin{figure}[hbtp]
    \centering
    \vspace{-10 pt}
    \includegraphics[width=0.8\linewidth, trim={0 0 0 0pt},clip]{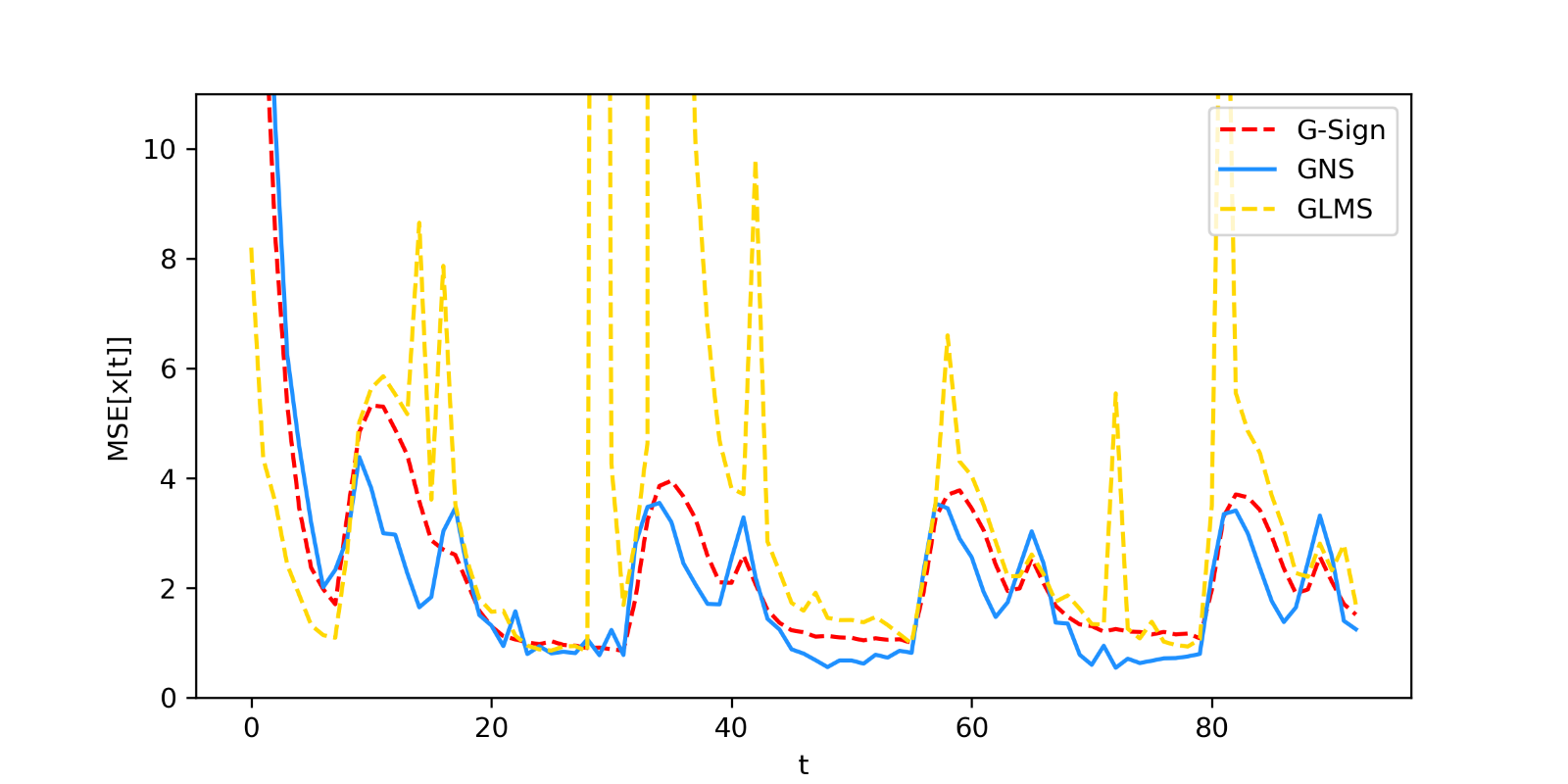}
    \vspace{-10 pt}
    \caption{Spatial MSE of GNS, G-Sign and GLMS.}  
    \vspace{-10 pt}
    \label{fig_convergence}
\end{figure}

% \section*{Acknowledgment}

% The preferred spelling of the word ``acknowledgment'' in America is without 
% an ``e'' after the ``g''. Avoid the stilted expression ``one of us (R. B. 
% G.) thanks $\ldots$''. Instead, try ``R. B. G. thanks$\ldots$''. Put sponsor 
% acknowledgments in the unnumbered footnote on the first page.

\end{document}